\documentclass{article}
\usepackage{graphicx}
\usepackage{epsfig}
\usepackage{amsmath}
\usepackage{amsfonts}
\usepackage[utf8]{inputenc}
\usepackage[T1]{fontenc}
 \usepackage{amssymb}
\usepackage[dvips]{color}
\date{\today}

\newcommand{\insertplot}[5]{\begin{figure}
 \hfill\hbox to 0.05in{\vbox to #5in{\vfill
 \inputplot{#1}{#4}{#5}}\hfill}
 \hfill\vspace{-.1in}
 \caption{#2}\label{#3}
 \end{figure}}
 \newcommand{\inputplot}[3]{
 \special{ps: plotfile #1}
}
\newcounter{fig}

\newcommand{\ee}{\end{equation}}
\newcommand{\eea}{\end{eqnarray}}
\newcommand{\be}{\begin{equation}}
\newcommand{\bea}{\begin{eqnarray}}

\begin{document}

 \title{ 
Charged black holes in a generalized scalar-tensor gravity model
} 

\author{
{\large Yves Brihaye}\\
{\small Physique Th\'eorique e Math\'ematiques, Universit\'e de
Mons, Place du Parc, 7000 Mons, Belgique}\\
\\
{\large Betti Hartmann}\\
{\small Instituto de F\'isica de S\~ao Carlos (IFSC), Universidade de S\~ao Paulo (USP), CP 369, 13560-970, S\~ao Carlos, SP, Brasil}
}
%%%
\maketitle 
\begin{abstract} 
We study 4-dimensional charged and static black holes in a generalized scalar-tensor gravity model, in which a shift symmetry for the scalar field 
exists. For vanishing scalar field the solution corresponds to the Reissner-Nordstr\"om (RN) solution, while solutions of the full scalar-gravity model
have to be constructed numerically. We demonstrate that these black holes support galilean scalar hair up to a maximal value of 
the scalar-tensor coupling that depends on the value of the charge and can be up to roughly  
twice as large as that for uncharged solutions. The Hawking temperature $T_{\rm H}$ of the hairy black holes at maximal scalar-tensor coupling 
decreases continuously
with the increase of the charge and reaches $T_{\rm H}=0$ for the highest possible charge that these solutions can carry.
However, in this limit, the scalar-tensor coupling needs to vanish. The limiting solution hence corresponds to the extremal RN solution, which does
not support regular galilaen scalar hair due to its AdS$_2\times S^2$ near-horizon geometry.

\end{abstract}

%%%%%%%%%%%%%%%%%%%%%%%%%%%%%%%%%%%%%%%%%%%%%%%%%%%%%%%%%%%%%%%%%%%%%%%%%%%%%%
\section{Introduction}
Black holes are amongst the most fascinating predictions of the best tested theory of gravity that we have to this day: General Relativity 
\cite{EinsteinGR}.
Already one year after Einstein formulated this theory, Schwarzschild found a static and spherically symmetric solution to the vacuum 
Einstein equation \cite{schwarzschild,chandra}
that was later on used to describe the exterior of an object so dense that for an observer far from it
light becomes infinitely red-shifted on the so-called {\it event horizon} of this object, which was successively called a {\it black hole}.
Interestingly, Schwarzschild's solution
is described only in terms of one single conserved quantity measured at infinity, namely the ADM mass \cite{ADM}.
Israel \cite{israel1} formulated a theorem named after him stating that the only 
static and asymptotically-flat vacuum space-time possessing a regular horizon is the Schwarzschild solution.

Charged as well as rotating generalizations were constructed since then \cite{reissner,kerr} and these are again
described by a small number of conserved quantities at infinity. The charged and static solution -- often referred to as the 
Reissner-Nordstr\"om (RN) 
solution -- is described by its charge as well as ADM mass, while the Kerr(-Newman) solution, which is a rotating solution of the (electro)vacuum 
Einstein equation, is described by its angular momentum (, charge) and ADM mass. Uniqueness and no-hair conjectures have been
formulated in this context \cite{israel2,carter,robinson,heusler,MTW}. It has since then been found that these conjectures do not hold
necessarily in the presence of non-linear matter sources such as Skyrme fields \cite{moss} or Yang-Mills fields \cite{bizon}.

The RN and Kerr(-Newman) solutions played a crucial r\^ole
in the discovery of Black hole Thermodynamics, which assigns properties to black holes alike those of thermodynamical systems, e.g
the Hawking temperature $T_{\rm H}$ \cite{hawking} which is proportional to the surface gravity of the black hole on its horizon or the
Bekenstein-Hawking entropy $S$ \cite{entropy} which is proportional to the horizon area. 
Moreover, it was shown that black holes fulfill laws that are equivalent to those of thermodynamics \cite{bh_thermo}. The third law of thermodynamics
states that as the temperature of the system tends to absolute zero, the entropy should also tend to zero. Black holes with
$T_{\rm H}=0$ do exist -- normally referred to as {\it extremal black holes} -- but these have non-vanishing horizon area and hence
non-vanishing Bekenstein-Hawking entropy. By counting the microstates of certain 5-dimensional extremal black holes, the expression of
Bekenstein-Hawking was confirmed to be proportional to the horizon area and hence non-vanishing \cite{strominger_vafa}. However,
in \cite{entropy_zero} it was suggested that the entropy of the extremal RN solution should be vanishing, despite its
non-vanishing horizon area. The essential reason for that is that the extremal horizon is an infinite proper distance away from 
any point outside the horizon. 

Extremal black holes typically
exist in systems where the (attractive) gravitational force is balanced by a (repulsive) force caused by electromagnetic fields and/or rotation.
Moreover, they possess a very interesting property: their near-horizon geometry is typically given by a product of a 2-dimensional
Anti-de Sitter geometry AdS$_2$ and another constant curvature manifold \cite{ads2_extremal} (see also \cite{Carroll} for a detailed
discussion on the geometry of extremal RN solutions). The fact, that an AdS factor appears makes it possible
to associate a dual Conformal Field Theory (CFT) to it via the AdS/CFT correspondence \cite{adscft} and hence compute the
black hole entropy via the dual CFT \cite{Strominger}. 

Besides possessing an extremal limit, the RN and Kerr solutions share another interesting feature: they possess two horizons,
one of which is the above mentioned event horizon, while the horizon lying inside the event horizon corresponds to a surface
of infinite blue-shift and is often referred to as a {\it Cauchy horizon}. This leads to the fact that the central physical curvature
singularity is time-like instead of space-like as in the case of the Schwarzschild solution. An observer crossing the Cauchy horizon can
hence avoid the physical singularity by travelling to another asympotically flat space-time region \cite{chandra}. To state it differently: the causal structure
of the Kerr space-time is much more alike that of the RN space-time. The latter is hence often used as a simpler spherically
symmetric, static analogue of the former \cite{relation_RN_Kerr}. We will adopt this view in the present letter as well and will study charged, static
and spherically symmetric black hole solutions also with the motivation to learn something about axially symmetric, stationary, rotating
solutions. 

While General Relativity provides are very accurate description of astrophysical and cosmological processes, the question remains
whether it is also applicable on microscopic scales, i.e. when quantum effects become important. General Relativity is now 
regarded as a low energy effective limit of a more general Quantum Theory of Gravity, which involves additional fields such as 
scalar fields and gauge fields. And since black holes are the ideal testing ground for such theories, 
it is important to understand how the presence of other fields alters the properties of black hole solutions.
Moreover, while black hole have been discussed extensively on theoretical grounds over the past decades, 
there exists no real ``proof'' for the existence of black holes up until now,
though the direct observation of gravitational waves seems to give strong evidence in favour of their existence \cite{gw1}. 
It is hence interesting to understand -- in particular with view to recent and future high precision astrophysical and cosmological data --
whether black holes can be used to test any Quantum Gravity Model.

In the following, we will concentrate on the addition of scalar fields to generalized gravity models. Scalar fields appear in low energy limits 
of String Theory (e.g. dilaton, axion) as generic fields, but are also used as effective description of physical systems, e.g. in the Ginzburg-Landau model
of superconductivity \cite{GL}.

A 4-dimensional low energy effective model of String Theory is the Einstein-Gauss-Bonnet-dilaton (EGBD) model. In this case,
the higher order curvature correction in terms of the Gauss-Bonet term ${\cal G}$ is coupled to the dilaton, a real scalar field $\phi$,
via a term of the form $\exp(\phi){\cal G}$. Static, spherically symmetric black hole solutions to this model have been first discussed in \cite{Kanti:1995vq}
and it was shown that black holes can carry a non-vanishing dilaton field, i.e. an example of ``scalar hair'' on the horizon (see also
\cite{Guo2008} for a review and \cite{Zang2015} for the discussion of the thermodynamics of black branes in this model).

In recent years, so-called {\it generalized scalar-tensor gravity} models in 4 space-time dimensions have been discussed. 
The idea goes back to the construction
of Horndeski \cite{horndeski} and has been extended in the past 10 years to so-called Galilaen gravity models \cite{Nicolis:2008in, Deffayet:2011gz} 
(for a review see also \cite{Deffayet:2013lga}) in which the scalar field possesses a shift symmetry (i.e. a symmetry under a Galilean transformation)
of the form $\phi\rightarrow \phi + c + a_{\mu}x^{\mu}$, where $c$ is a constant and $a_{\mu}$ a constant co-vector. 
This symmetry has a conserved Noether current $J^{\mu}$ associated to it:~$\nabla_{\mu} J^{\mu}=0$.

A no-hair conjecture for black holes in generic Galilean theories has been presented in \cite{Hui:2012qt}, however,
for specific choices of the Galilean action it has been shown that black holes with scalar hair can be constructed (numerically) 
\cite{Sotiriou:2013qea,Sotiriou:2014pfa}. In these models, the scalar field couples directly to the Gauss-Bonnet term in the form $\phi{\cal G}$. This leads to an equation of motion
for the scalar field in which the massless and real scalar field is sourced by the Gauss-Bonnet term and coupled to it via the scalar-tensor coupling. 
Note that black holes with galilean scalar hair can also be constructed  in models without this specific coupling between the scalar field
and the Gauss-Bonnet term \cite{Babichev2013,Babichev2016,Babichev2017} and that this construction can be extended to include charge \cite{Babichev2015}.
In \cite{Babichev2017} it was also realized that the black holes constructed in \cite{Sotiriou:2014pfa}
evade the no-hair conjecture of \cite{Hui:2012qt} because the norm of the Noether current diverges at the horizon of the black hole. However, one of the key assumption of \cite{Hui:2012qt} is that the norm of the Noether current
be finite on the horizon and it was further demonstrated in \cite{Babichev2017} that with the assumption
of vanishing radial Noether current, black hole solutions do not exist in these class of models. However, modifying the scalar-tensor gravity model allows for 
black hole solutions with finite norm of the Noether current \cite{Lehebel:2017fag}.

In this letter we study charged, static, spherically symmetric black holes in the generalized scalar-tensor gravity model presented in 
\cite{Sotiriou:2014pfa}. For vanishing scalar-tensor coupling the solution corresponds to the RN solution.
We report on a first study of these solutions and focus on the construction of the solutions outside the event horizon.
In particular we discuss the extremal limit of these solutions and show that they do not support scalar galilean hair. 

Our letter is organized as follows: in Section 2, we give the model, the Ansatz and the boundary conditions. In Section 3, we discuss our analytic 
as well as numerical results, while Section 4 contains our conclusions.

%%%%%%%%%%%%%%%%%%%%%%%%%%%%%%%%%%%%%%%%%%%%%%%%%%%%%%%%%%%%%%%%%%%%%%%%%%%%%%%  
  
%%%%%%%%%%%%%%%%%%%%%%%%%%%%%%%%%%%%%%%%%%%%%%%%%%%%%%%%%%%%%%%%%%%%%%%%%%%%%%%
\section{The model}
%%%%%%%%%%%%%%%%%%%%%%%%%%%%%%%%%%%%%%%%%%%%%%%%%%%%%%%%%%%%%%%%%%%%%%%%%%%%%%% 

%%%%%%%%%%%%%%%%%%%%%%%%%%%%%%%%%%%%%%%%%%%%%%%%%%%%%%%%%%%%%%%%%%%%%%%%%%%%%%%

%%%%%%%%%%%%%%%%%%%%%%%%%%%%%%%%%%%%%%%%%%%%%%%%%%%%%%%%%%%%%%%%%%%%%%%%%%%%%%% 
The scalar-tensor gravity model we are working with here has been studied in detail without additional matter content \cite{Sotiriou:2014pfa}. The gravity part of
this model is invariant under shift of the scalar field leading to an associated conserved Noether current.
In this letter, we couple the model to a U(1) gauge field. The action then reads 

\be
\label{action}
S =  \int  {\rm d}^4x  \sqrt{-g} \left[\frac{R}{16\pi G}  +  \frac{\gamma}{2}  \phi {\cal G} -  \frac{\beta}{2}  
\partial_{\mu} \phi  \partial^{\mu} \phi  - \frac{1}{4} F_{\mu\nu}F^{\mu\nu} \right] \ ,
\ee
where the Gauss-Bonnet term ${\cal G}$ and the field strength tensor $F_{\mu\nu}$ of the U(1) gauge field $A_{\mu}$ are given by
\be
 {\cal G} = R^{\mu \nu \rho \sigma} R_{\mu \nu \rho \sigma} - 4 R^{\mu \nu}R_{\mu \nu} + R^2  \ \ , \ \ F_{\mu\nu}=\partial_{\mu} A_ {\nu}- \partial_{\nu} A_{\mu} \ ,
\ee
respectively.  $\gamma$ and $\beta$ are dimensional coupling constants. In the case $\phi\equiv const.$ the Gauss-Bonnet term
becomes a total divergence and we are left with the standard Einstein-Hilbert action coupled minimally to a U(1) gauge field. 
Appropriate scalings of the coordinates and fields allow us to set $\beta^{-1}=8\pi G= 1$ such that the action depends only on the parameter $\gamma$. 
Varying the  action (\ref{action}) with respect to the metric, the scalar field and the gauge field, we obtain the following set of equations:
\begin{equation}
\label{eom}
\square \phi = - \frac{\gamma}{2} {\cal G} \  \  \  \   ,  \  \ \ \  \partial_{\mu} \left(\sqrt{-g} F^{\mu\nu}\right)=0 \ \ \ \  , \  \ \ \ 
G_{\mu\nu} - \partial_{\mu} \phi \partial_{\nu} \phi + \frac{1}{2}g_{\mu\nu} \partial_{\alpha} \phi \partial^{\alpha} \phi + \frac{\gamma}{2}
{\cal K}_{\mu\nu} = T_{\mu\nu}^{\rm (EM)} \ ,
\end{equation}
where 
\be
       {\cal K}_{\mu\nu} =( g_{\rho\mu} g_{\sigma\nu} + g_{\rho\nu} g_{\sigma\mu})\nabla_{\lambda} (\partial_{\gamma} \phi \epsilon^{\gamma \sigma\alpha\beta} \epsilon^{\delta\eta } R_{\delta\eta\alpha\beta})
\ee
results from the variation of the Gauss-Bonnet term with respect to the metric and
$T_{\mu\nu}^{\rm (EM)} $ is the energy-momentum tensor of the gauge field
\be
T_{\mu\nu}^{\rm (EM)} =  F_{\mu\alpha}F_{\nu}^{\alpha} - \frac{1}{4} g_{\mu\nu} F_{\alpha\beta}F^{\alpha\beta} \  .
\ee
The locally conserved Noether current associated to the shift symmetry reads \cite{Babichev2017}~:
\begin{equation}
J^{\mu} = \frac{1}{\sqrt{-g}} \frac{\delta S[\phi]}{\delta (\partial_{\mu} \phi)} \ .
\end{equation}

\subsection{Ansatz and boundary conditions}
We are studying spherically symmetric, static, electrically charged black hole solutions in this letter. The Ansatz for the metric, scalar field and gauge field,
respectively, reads
\begin{equation}
ds^2=-N(r) A(r)^2 dt^2 + \frac{1}{N(r)} dr^2  + r^2 \left(d\theta^2 + \sin^2 \theta d \varphi^2\right)  \ , \   \phi=\phi(r) \ , \ 
A_{\mu} dx^{\mu} = V(r) dt  \ .
\end{equation}
Inserting this Ansatz into the equations of motion (\ref{eom}) results in a coupled system of ordinary differential equations that has to be solved
subject to the appropriate boundary conditions. Asymptotic flatness and finite energy requires
\begin{equation}
\label{bcinf}
  A(r \to \infty) = 1 \ \ , \ \ \left(r^2 V'\right)\bigg\rvert_ {r \to \infty} = Q \ , 
\end{equation}
where $Q$ is the charge of the solution (in appropriate units) and the prime now and in the following denotes the derivative with respect to $r$. 

We can use the shift symmetry of the scalar field and the fact that the electric potential is determined only up to a constant to fix
the scalar field and the potential $V$ on the horizon to 
\be
       \phi(r_h)=0 \ \ , \ \ V(r_h) =0  \ .
\ee
Note that we could have equally fixed the value of the fields at infinity, but we found that our numerical results are more accurate with the choice
of the fields on the horizon.
Furthermore, the requirement of regularity of the fields on the horizon $r=r_h$  implies $N(r_h)=0$
as well as the following constraint:
\begin{equation}
\label{condition}
\Bigl[\gamma (V')^4 r^3(2 \gamma \phi'-r) 
+ 4 A^2 (V')^2 \Bigl( \gamma(2 \gamma^2+r^4) (\phi')^2 + (2 \gamma^2 +r^4)r \phi' + 6 \gamma r^2 \Bigr)
- 16 A^4 \Bigl(\gamma r^2 (\phi')^2   + r^3 \phi' + 3\gamma \Bigr)\Bigr] \bigg\rvert_{r=r_h} = 0 \ .
\end{equation}
This condition can be simplified by using
the Maxwell equation which implies a relation between the potential $V(r)$ and the metric function $A(r)$~:
\begin{equation}
\label{relation}
\frac{dV(r)}{dr}=c\frac{A(r)}{r^2} \ ,
\end{equation}
where $c$ is an integration constant, which is fixed by the boundary conditions (\ref{bcinf}) to be $c=Q$.
Using the above relation at $r=r_h$, the condition  (\ref{condition})  becomes independent of $A(r_h)$ and
can  be solved for $\phi'$. We find~:
\begin{equation}
\label{phip}
\phi'_{\pm}\bigg\rvert_{r=r_h}=-\frac{\gamma^2 Q^4 + 4\gamma^2Q^2r_h^2 + 2Q^2 r_h^6 - 8 r_h^8 \pm \sqrt{\vert Q^2 - 4 r_h^4\vert} \sqrt{\Delta_1}}
{4\gamma r_h(2\gamma^2 Q^2 + Q^2 r_h^4 - 4 r_h^6)} 
\end{equation}
with
\begin{equation}
\label{delta1}
\Delta_1=\gamma^4 Q^6 + 4\gamma^2  r_h^2 Q^4 (5 \gamma^2 + 2 r_h^2) 
- 4 Q^2 r_h^2(24 \gamma^4 + 20 \gamma^2 r_h^4 - r_h^8) 
+ 16 r_h^{10}(12 \gamma^2 - r_h^4)  \ .
\end{equation}
The equation $\Delta_1=0$ then gives the critical value of $\gamma$ in dependence on $Q$ and $r_h$. Moreover, $\phi'$ diverges 
on the horizon for $Q^2\rightarrow 4r_h^6/(2\gamma^2 +r_h^4)$.

\subsection{Physical properties of black holes}
The  temperature $T_{\rm H}$ and the entropy $S$ of the black hole in rescaled coordinates and with our Ansatz read, respectively~:
\begin{equation}
T_{\rm H}=\frac{1}{4\pi} \left(N'A\right)_{r=r_h} \  \ , \ \  S=\frac{{\cal A}_h}{4}=\pi r_h^2 \ ,
\end{equation}
with ${\cal A}_h$ denoting the horizon area. The ADM mass can be read off from the behaviour of the metric function $N(r)$ at spatial infinity. Definining
the mass function $m(r)$ via $N(r)=1-2m(r)/r$, the ADM mass $M$ (in rescaled units) reads
\begin{equation}
M=\lim_{r\rightarrow \infty} m(r) \ ,
\end{equation}
while the charge $Q$ is determined by the asymptotic behaviour of $V(r)$, see (\ref{bcinf}). 

The radial component of the conserved Noether current associated to the shift symmetry reads \cite{Babichev2017}~:
\begin{equation}
J^r = N\left[ \frac{\gamma (N-1)}{2 r^2} \left(\frac{N'}{N} + \frac{2A'}{A} \right) - \phi'\right] \ 
\end{equation}
and the norm of the current will be
\begin{equation}
\sqrt{J_r J^r} =  \frac{\gamma (N-1)}{2 r^2} \left(\frac{N'}{N} + \frac{2A'}{A} \right) - \phi'  \ .
\end{equation}
Using that $N(r_h)=0$ and $N'\vert_{r_h}\neq 0$ for non-extremal black holes (see the discussion below for the extremal case), it is
easy to see that the norm of the radial Noether current diverges on the horizon of the solutions constructed in this paper. We, in fact, checked numerically that the radial
component of the Noether current $J^r$ is non-vanishing for all black hole solutions that carry scalar hair. Hence, very similar to the uncharged case, our solutions
evade the No-hair theorem of \cite{Hui:2012qt} by violating the assumption of finite
norm of the Noether current on the horizon.

\section{Charged black holes}
As mentioned above, the equations allow an explicit solution in the case $\gamma=0$. This is the RN solution
\cite{chandra,reissner}:
\be
\label{RN}
       N(r) = 1 - \frac{2 M}{r} + \frac{Q^2}{4 r^2} \ \ , \ \  V(r)=\frac{Q}{r_h}-\frac{Q}{r} \ \ , \ \ A(r)\equiv 1
\ee
with $M$ and $Q$ the mass and charge of the solution, respectively. This solution has two horizons at $r_{\pm}=M\pm \sqrt{M^2- Q^2/4}$. $r_+$ corresponds
to the event horizon, while $r_-$ is the Cauchy horizon. The extremal RN has $M=Q/2$ such that $r_+=r_-$ and $T_{\rm H}\rightarrow 0$, while
the entropy $S$ stays finite. The Kretschmann invariant $K=R_{\mu\nu\rho\sigma} R^{\mu\nu\rho\sigma}$ is perfectly finite
 at $r=r_{\pm}$. It only diverges at $r\rightarrow 0$, i.e. at the physical singularity of the space-time. 

For $\gamma > 0$, the corresponding equations do not admit explicit solutions, so we need to solve the equations numerically. We have used
a grid adaptive collocation solver \cite{COLSYS} to find solutions to the system of coupled, non-linear ordinary differential equations. Our results 
are described in the following section.

\subsection{Charged black holes with scalar hair}

Now and in the following, we will use the fact that we can rescale the radial variable $r$ 
to set $r_+=r_h=1$. The equation $\Delta_1=0$ (see (\ref{condition})) then gives the critical value $\gamma=\gamma_c$ for a fixed $Q$:
\begin{equation}
\label{gammacr}
{\gamma}^{(\pm)}_{c} = \sqrt{\frac{\left(24 - 4 Q^2\right)\pm 2\sqrt{3(Q^4 - 24Q^2 + 48)}}{Q^2(Q^2 + 24)} } 
\end{equation}
or the critical value of the charge $Q=Q_c$ for a fixed value of $\gamma$:
\begin{equation}
\label{qcr}
 Q_c=\frac{\sqrt{2}}{\gamma}
 \sqrt{\left(-6\gamma^4 - 2 \gamma^2 + \gamma\sqrt{3(12\gamma^4 + 12\gamma^2+1)} \right)}
\end{equation}
As discussed in detail in {\cite{Sotiriou:2014pfa} $\Delta_1=0$ gives $\gamma_c= 1/\sqrt{12} \approx 0.289 $ for $Q=0$.
For $\gamma=0$ it follows from (\ref{condition}) that $Q=2$. Note that $Q\rightarrow 2$ for $\gamma\rightarrow 0$ leads to $\phi'\vert_{r=r_h}\rightarrow -\infty$.

We have integrated the equations of motion for $r \ge 1$, i.e. outside the horizon $r_h=1$.
Our numerical results suggest that there are three different regimes in the $\gamma$-$Q$-plane that limit the existence of solutions:
\begin{enumerate}
 \item $Q \ll 1$: this includes the $Q=0$ case studied in (\cite{Sotiriou:2014pfa}) for which $\gamma_c\approx 0.289$. Increasing the charge 
 $Q$ from this solution, we find a branch of charged black holes that carry galilaen scalar hair on their horizon. Since $Q$ enters (\ref{delta1}) at order O$(Q^2)$,
 we expect that -- as long as $Q$ is small -- the value of $\gamma_c$ not to differ much from its value at $Q=0$. This is confirmed by our numerics,
 see Fig.\ref{fig1}, where we give the critical value of $Q_c$ in dependence on $\gamma$ (or vice versa, $\gamma_c$ in terms of $Q$) for charged hairy black hole
 solutions. We also show the Hawking temperature $T_{\rm H}$ in dependence on $\gamma$. The uncharged solution is the
 solution with highest temperature and the increase of the charge $Q$ lowers the temperature.
 Some physical properties of the uncharged solution ($Q=0$) are shown in Fig.\ref{fig2}. In this case, the value of $\phi'$ at $r=r_h$ is
 (see (\ref{phip})): $\phi'_{\pm}(r_h)=(-1\pm \sqrt{1-12\gamma^2})/(2\gamma)$. Increasing $\gamma$ from zero, the derivative
 of $\phi'$ decreases on the $\phi'_+(r_h)$ branch until it reaches its critical value at $\gamma=\gamma_c=1/\sqrt{12}$ which 
 is $\phi'_{+,c}(r_h)=-\sqrt{12}/2\approx -1.732$. Our numerical results indicate that from this value of $\gamma$, a second branch extends backwards in
 $\gamma$ corresponding to values of $\phi'_{-}(r_h)$ such that $\phi'(r_h)$ further decreases. Since the objective of this
 paper has not been to study the uncharged case newly, we have not investigated this in more detail, but we remark that our numerical
 results indicate that this second branch exists though it has not been mentioned in the literature previously.

 \begin{figure}
\begin{center}
{\includegraphics[width=8cm]{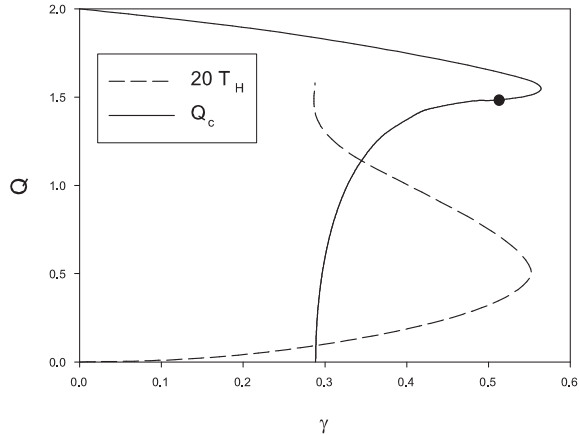}}
\caption{We show the critical value of the charge $Q_c$, up to where charged black holes with galilaen scalar hair exist, in dependence
on the parameter $\gamma$ (solid). We also give the Hawking temperature $T_{\rm H}$ of the black holes with 
charge $Q=Q_c$ in dependence on $\gamma$ (dashed). The dot at $Q\approx 1.484$ and $\gamma\approx 0.513$ corresponds to the point at which
$Q_c^4 -24Q_c^2 +48=0$ (see discussion in the text).
\label{fig1}}
\end{center}
\end{figure}
 
 \begin{figure}
\begin{center}
{\label{fig2a}\includegraphics[width=8cm]{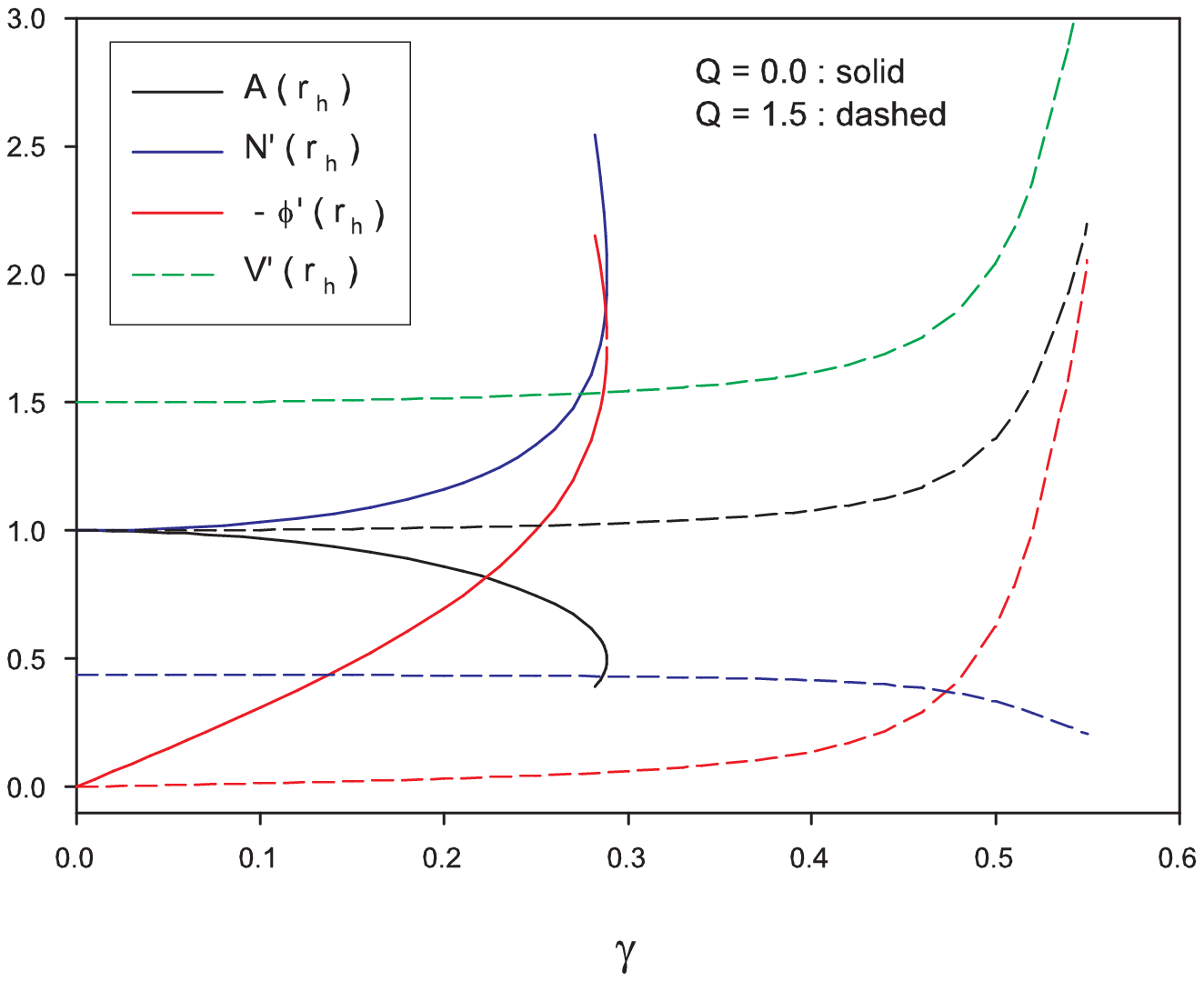}}
{\label{fig2b}\includegraphics[width=8cm]{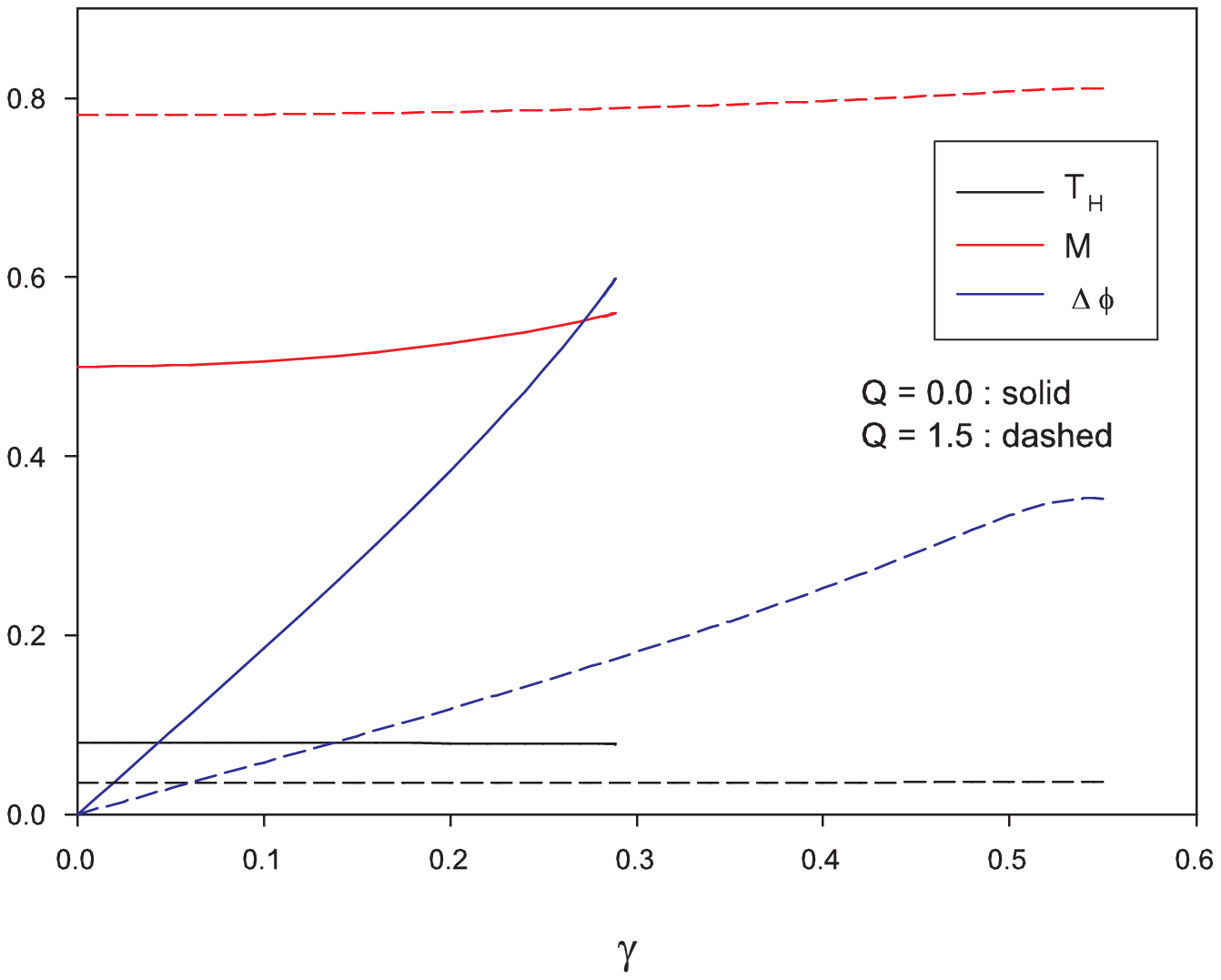}}
\caption{ {\it Left:} We show the value of the metric function $A(r)$ (black) as well as the derivatives $N'(r)$ (blue), $\phi'(r)$ (red), $V'(r)$ 
(green) of the metric function $N(r)$, 
the scalar field $\phi(r)$ and the electric potential $V(r)$, respectively, at the horizon $r_h$ in dependence on $\gamma$ for
$Q=0$ (solid) and $Q=1.5$ (dashed). 
{\it Right :} We show the Hawking temperature $T_{\rm H}$ (black), the ADM mass $M$ (red) and $\Delta\phi=\phi(\infty)-\phi(r_h)$ (blue) in dependence on $\gamma$ 
for $Q=0$ (solid) and $Q=1.5$ (dashed).
\label{fig2}}
\end{center}
\end{figure}

\begin{figure}
\begin{center}
{\includegraphics[width=8cm]{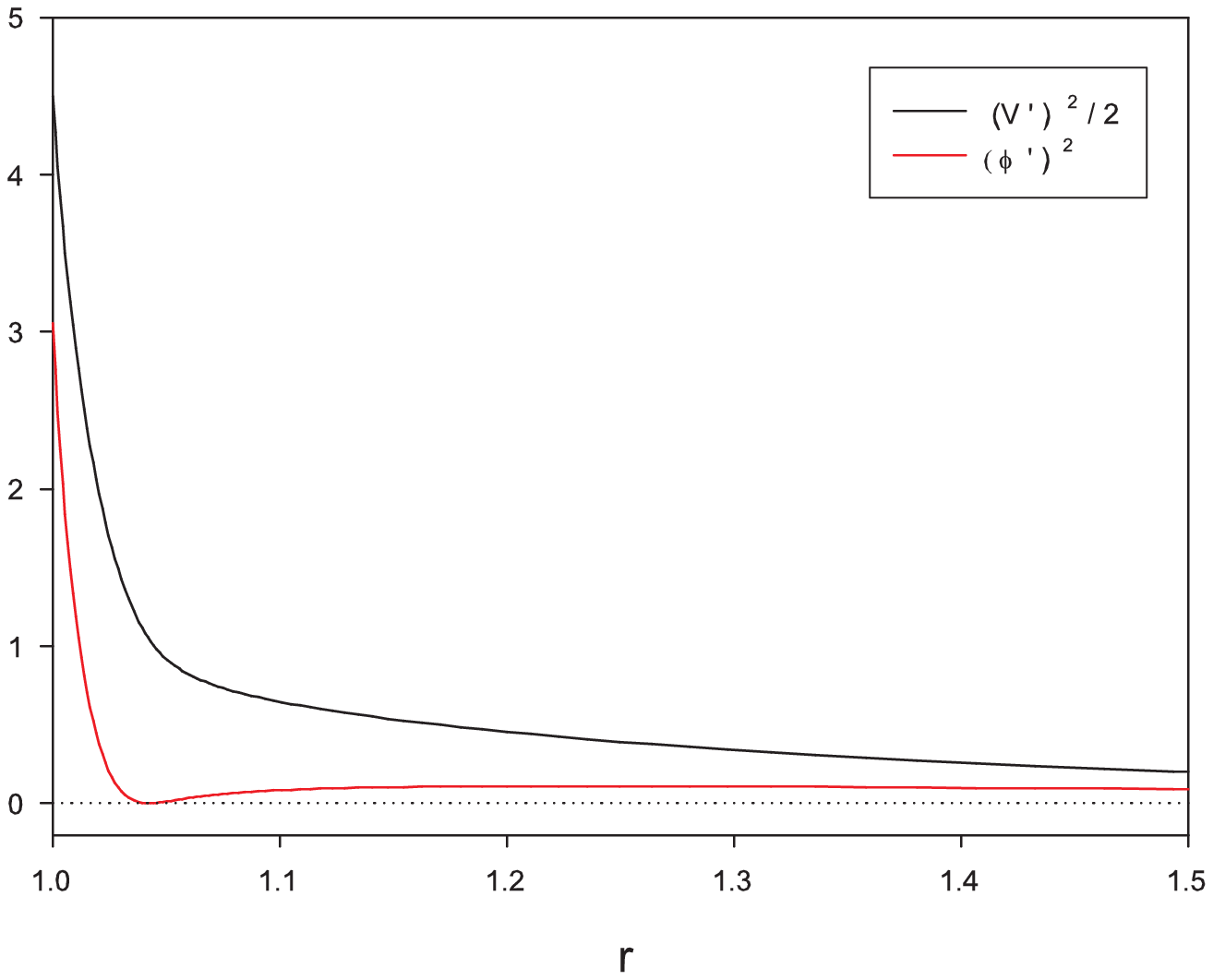}}
{\includegraphics[width=8cm]{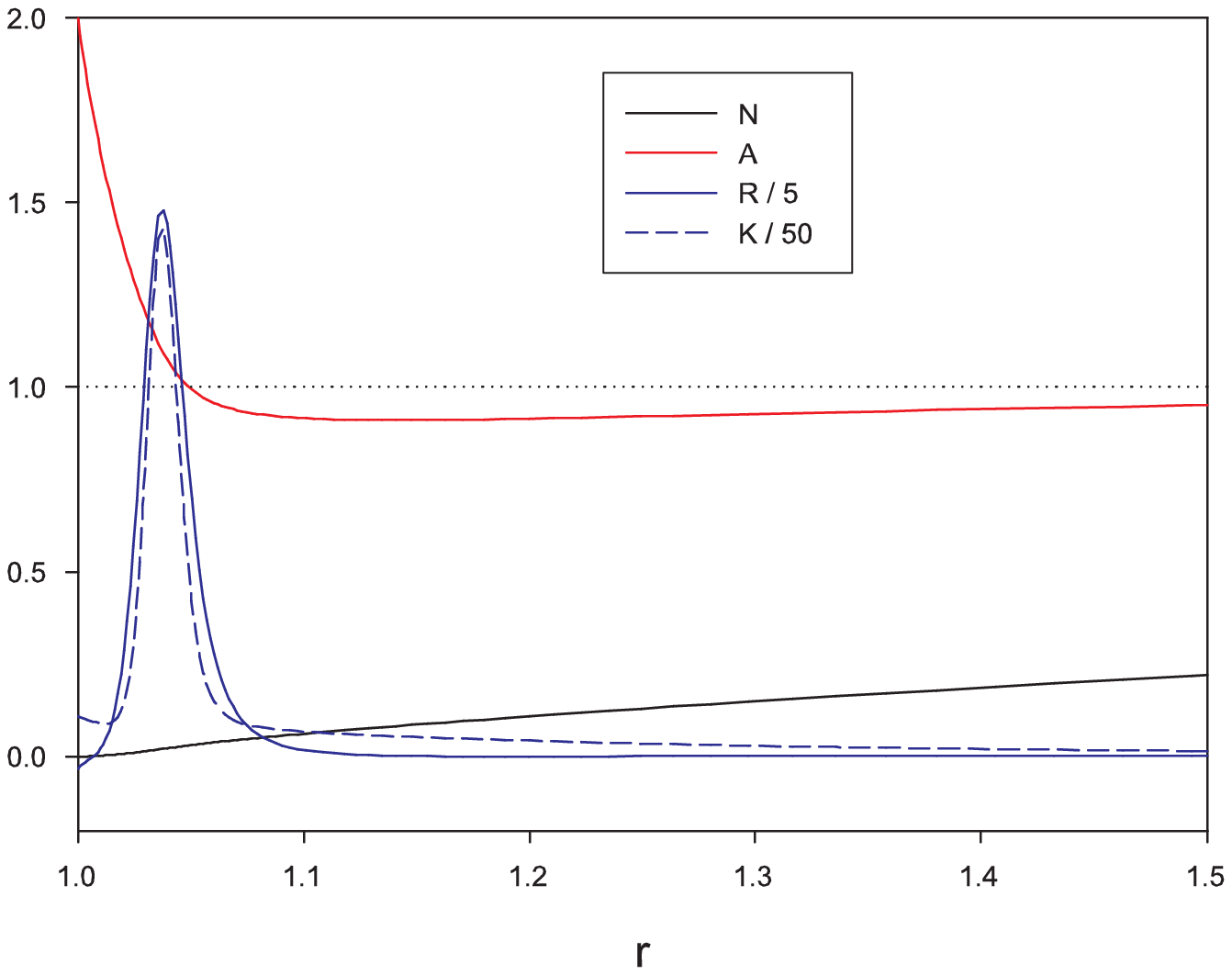}}
\caption{ {\it Left :} 
We show $(\phi')^2$ and $(V')^2$ in function of the radial coordinate $r$ close to $r_h=1$ for $Q\approx Q_M=1.5$ and $\gamma=0.55$ close to $\gamma_M \approx 0.56$. 
{\it Right :} We show the metric functions $N(r)$ and $A(r)$ as well as the Ricci scalar $R$ and the Kretschmann scalar $K=R_{\mu\nu\rho\sigma} 
R^{\mu\nu\rho\sigma}$ in function of the radial coordinate $r$ close to $r_h=1$ for $Q\approx Q_M=1.5$ and $\gamma=0.55$ close to $\gamma_M \approx 0.56$. 
\label{fig3}
}
\end{center}
\end{figure} 
 
 \item $Q \approx 1$: when $Q$ becomes of the order of unity, we find that $\gamma_c$ strongly increases. 
 The critical values of $Q$ and $\gamma$, respectively, are
 still determined by  $\Delta_1=0$. The condition, $\Delta_1=0$, however can only be fulfilled up to a
 certain value of the charge $Q$. The limiting critical value of the charge $Q_c=\tilde{Q}_c$ is given
 by the condition $\tilde{Q}_c^4 -24\tilde{Q}_c^2 +48=0$ (see \ref{gammacr}). 
 This gives $\tilde{Q}_c= \left(12 - 3\sqrt{6}\right)^{1/4} \approx 1.484$ with corresponding value $\tilde{\gamma}_c=\sqrt{(-2\sqrt{6}+3)/(24\sqrt{6}-66)}
 \approx  0.513$. Our numerical
 results confirm this reasoning, see Fig.\ref{fig1}, in which we have marked this point by a dot.
 We also observe that the temperature of the black hole decreases continuously with the increase of the charge.

 \item $Q \gtrsim \tilde{Q}_c\approx 1.484$: increasing the charge $Q$ further, $\Delta_1$ is always positive and no critical behaviour results from (\ref{delta1}).
 We find that first the value of $\gamma_c$ increases when increasing $Q$ up to a maximal value $\gamma_M\approx 0.56$ and corresponding 
 $Q=Q_M \approx 1.5$.  
 Examining the  solutions for $\gamma\in [\tilde{\gamma_c}:\gamma_M]$ further shows that the values of $\phi'$, $V'$ and $A$ on the horizon
 are large, see Fig.\ref{fig3} (left) for $\gamma=0.55$, but that there exists a finite radius surface, 
 which lies outside the horizon on which both the Kretschmann scalar $K$ as well as the Ricci scalar $R$ become very large (see Fig.\ref{fig3} (right)).
 The metric function $N(r)$ stays perfectly finite in this case. Note that the location of this surface is given by the double zero of 
 the derivative of the scalar field function $(\phi')^2$, see Fig.\ref{fig3} (left). 
 
 In Fig. \ref{fig2} (left) we show the values of the metric and matter fields at the horizon in dependence on $\gamma$ for $Q=1.5$. This figure show clearly
 the difference between the charged and uncharged case, in particular we note that there is no back-bending of the curves for $Q=1.5$. 
 Moreover, the value $N'(r_h)$ is continuously increasing in the uncharged case, while it decreases in the charged case. 
 The right side of Fig. \ref{fig2} demonstrates that the Hawking temperature for the charged black hole is smaller than that of the uncharged one, the 
 ADM mass is larger, and the difference $\Delta\phi:=\phi(\infty)-\phi(r_h)$, which is a physical quantity, increases slower than for the uncharged case.
 
 Note that $\gamma=\gamma_M\approx 0.56$ is the maximal possible scalar-tensor coupling for charged black hole, which is roughly twice as 
 large as that for uncharged solutions.

 However, increasing $Q$ further from $Q_M\approx 1.5$, $\gamma_c$ decreases from $\gamma_M$. In this case, we find that
 the phenomenon that limits the existence of solutions is very different from that for smaller charges. 
 We find that for this part of the branch  the denominator in (\ref{phip}) tends to zero indicating that the scalar field function
 has a diverging derivative at $r_h$. This happens for 
 \begin{equation}
  \gamma_c=\sqrt{\frac{2}{Q^2}-\frac{1}{2}} \ \ , \ \ Q_c=\sqrt{\frac{4}{2\gamma^2 +1}} \ .
  \end{equation}
  $\gamma_c$ hence becomes a decreasing function in terms of $Q$ and tends to zero for $Q\rightarrow 2$. Our numerical results
  confirm this, see Fig.\ref{fig1}. In Fig.\ref{fig4} we demonstrate this approach by showing the dependence of the ADM mass $M$, of the Hawking temperature $T_{\rm H}$, 
  the value of the scalar field derivative at the horizon $\phi'(r_h)$ and the physical difference $\Delta\phi=\phi(\infty)-\phi(r_h)$ 
  on the charge $Q$ for $\gamma=0.27$. While the ADM mass and the Hawking temperature show a very similar dependence on the charge $Q$
  as for the RN solution, we find that for $\gamma > 0$ and $Q$ sufficiently large that as $Q$ approaches $Q_c$ the derivative of 
  the scalar field function at the horizon diverges, i.e. $\phi'(r_h)\rightarrow -\infty$ for $Q\rightarrow Q_c$. For $\gamma=0.27$ we find that
  $Q_c\approx 1.85$.

  \begin{figure}
\begin{center}
{\includegraphics[width=8cm]{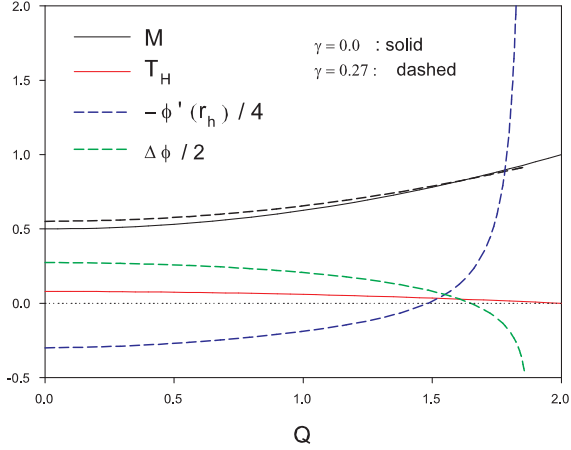}}
\caption{We show the dependence of the ADM mass $M$, of the Hawking temperature $T_{\rm H}$, the value of the scalar field
derivative at the horizon $\phi'(r_h)$ and the physical difference $\Delta\phi=\phi(\infty)-\phi(r_h)$ on the charge $Q$ for the 
RN
solution ($\gamma = 0$, solid) and   a charged black hole solution with scalar hair ($\gamma=0.27$, dashed).
Note that the two curves for $T_{\rm H}$ are indistinguishable.  \label{charge_vary} 
\label{fig4}}
\end{center}
\end{figure} 
  
  Note that the $\gamma=0$ limit corresponds to the RN solution, and to be more precise, the extremal RN solution, since
  $r_h=1=Q/2$. This is also indicated by the fact that the Hawking temperature $T_{\rm H}$ tends to zero in this limit.
  We hence find that {\it the solution with maximal possible charge $Q$ is the extremal RN solution} which does 
  not allow regular galilean hair on its horizon.  
\end{enumerate}

\subsection{No galilaen hair on extremal black holes}
\label{extremal}
Extremal black holes do usually not allow scalar field in their near-horizon geometry.
In the following we will demonstrate that this is also true for galilean scalar hair. For that we use the metric of 
an AdS$_2\times S^2$ space-time given by
\begin{equation}
\label{metric_ex}
ds^2=  v_1 \left(-\rho^2 d\tau^2 + \frac{1}{\rho^2} d\rho^2\right) + v_2 \left(d\theta^2 + \sin^2\theta d\varphi^2 \right)
\end{equation}
where $v_1$ and $v_2$ are positive constants. The coordinate $\rho$ is related to the coordinate $r$ by $\rho=r - r_h$ such that
$\rho\rightarrow 0$ corresponds to $r\rightarrow r_h$. Now considering the scalar field equation in this space-time, the equation $\square \phi = - \frac{\gamma}{2} {\cal G}$ and its solution read~:
\begin{equation}
\left(\rho^2 \phi'\right)'= 4\frac{\gamma}{v_2}  \ \ \ \Rightarrow \ \ \   \phi(\rho)=\frac{4\gamma}{v_2} \ln(\rho) + \frac{c_1}{\rho} + c_2  \ ,
\end{equation}
 where $c_1$ and $c_2$ are integration constants. This indicates that the scalar field and all its derivatives diverge at the horizon
 of an extremal black hole with $\rho\rightarrow 0$. 
 
 In Fig.\ref{fig4} we show the dependence of the ADM mass, the Hawking temperature $T_{\rm H}$, the value of $\phi'$ on the horizon
 as well as $\Delta\phi=\phi(\infty)-\phi(r_h)$  on the charge $Q$ for the RN solution ($\gamma=0$) and for a charged black hole
 solution with scalar hair. We have chosen $\gamma=0.27$ is this latter case such that black holes with scalar hair exist in the uncharged case $Q=0$.
 When increasing $Q$ to its maximal possible value $Q=2$ the two curves for the ADM mass and the Hawking temperature $T_{\rm H}$ join, indicating
 that the limiting solution at $Q=2$ with $T_{\rm H}=0$ does not carry scalar hair. Following our analytical arguments above as well
 as comparing with the numerical data show in Fig. \ref{fig4}, it is obvious that in this limit, the scalar field as well as its derivative
 on the horizon tend to infinity. Black holes regular at the horizon can only exist if $\gamma=0$, i.e. when the solutions
 are RN black holes, in this case the extremal solution with AdS$_2\times S^2$ horizon geometry.

\section{Conclusions and Outlook}
We have discussed charged black holes in a generalized scalar-tensor gravity model in which the scalar field is invariant
under a scale transformation. In this model, the Gauss-Bonnet curvature invariant sources the scalar field.
We demonstrate that charged black holes with scalar hair on their horizon exist, but that the existence is limited by a maximal value of the 
scalar-tensor coupling that depends on the value of the charge. We find that the maximal possible value of the 
scalar-tensor coupling for any charge is roughly twice as large as that for vacuum black holes of the model. 
Let us remark that although the charged black hole solutions with scalar hair have to be constructed numerically,
their domain of existence in the $\gamma$-$Q$-plane is (mainly) determined by analytical expressions, which agree to high accuracy with
our numerical results. 

The difference of our present study in comparison to the one done in \cite{Sotiriou:2014pfa} is the fact that extremal
black hole solutions do exist. Since extremal black holes are supposed to be important in the understanding of Black Hole Thermodynamics,
i.e. semiclassical models including gravity, as well as possible candidates for a full Quantum Theory of gravity such as String Theory, we have
studied the extremal limit in this letter as well. We come to the conclusion that the generalized scalar-tensor gravity model at hand
does not allow extremal solutions that carry scalar hair. 
Since the RN solution has a similar causal structure as the Kerr solution, we believe that similar arguments will hold in the case
of rapidly rotating black hole solution in scalar-tensor gravity when exchanging $Q$ by the angular momentum $J$ (for slow rotation see \cite{Sotiriou:2014pfa}). 

Finally, let us mention that we have only investigated the solutions outside of the event horizon. Following the analysis in \cite{Sotiriou:2014pfa}
one could also investigate the solution inside the event horizon of the black hole. Now, since the RN solution does
possess an additional Cauchy horizon that lies inside the event horizon, it will be interesting to see how the presence
of the scalar field changes the causal structure. In particular, it will be interesting to see how the location of
the finite surface on which the curvature invariants become very large (and probably diverge) is influenced. This is currently under investigation.

\vspace{1cm}

{\bf{Acknowledgements}} 
BH would like to thank FAPESP for financial support under grant number {\it 2016/12605-2}, CNPq for financial support
under {\it Bolsa de Produtividade Grant 304100/2015-3}, FAPES for partial financial support under contract number {\it 0447/2015} as well as
the excellence cluster EXC 1098 PRISMA for financial support. YB and (in particular) BH would like to thank the Mainz Institute for Theoretical Physics (MITP), and especially {\it Prof. Dr. Gabriele Honecker}, for the hospitality during their stay at the MITP in January 2017. We would also like to thank the anonymous referee of our paper for very valuable comments.

%%%%%%%%%%%%%%%%%%%%%%%%%%%%%%%%%%%%%%%%%%%%%%%%%%%%%%%%%%%%

%%%%%%%%%%%%%%%%%%%%%%%%%%%%%%%%%%%%%%%%%%%%%%%%%%%%%%%%%%%%

 \end{document}